\documentclass[conference]{IEEEtran}

\makeatletter

\def\ps@IEEEtitlepagestyle{%
  \def\@oddfoot{\mycopyrightnotice}%
  \def\@evenfoot{}%
}
\def\mycopyrightnotice{%
  {\footnotesize 979-8-3315-3559-9/25/\$31.00~\copyright~2025 IEEE\hfill}%
  \gdef\mycopyrightnotice{}
}

\usepackage{blindtext}
\usepackage{url}
\usepackage{eso-pic}
\IEEEoverridecommandlockouts
\usepackage{cite}
\usepackage{amsmath,amssymb,amsfonts}
\usepackage{algorithmic}
\usepackage{graphicx}
\usepackage{textcomp}
\usepackage{xcolor}
\def\BibTeX{{\rm B\kern-.05em{\sc i\kern-.025em b}\kern-.08em
    T\kern-.1667em\lower.7ex\hbox{E}\kern-.125emX}}    
\usepackage{eso-pic}
\usepackage{hyperref}
\newcommand\AtPageUpperMyright[1]{\AtPageUpperLeft{%
 \put(\LenToUnit{0.17\paperwidth},\LenToUnit{-2cm}){%
     \parbox{0.9\textwidth}{\raggedleft\fontsize{8}{11}\selectfont #1}}%
 }}%
\newcommand{\conf}[1]{%
\AddToShipoutPictureBG*{%
\AtPageUpperMyright{#1}
}
}

\begin{document}
\title{\vspace*{1cm} ``Test, Build, Deploy" - A CI/CD Framework for Open-Source Hardware Designs\\\thanks{This work was supported by NSF Award 171858}}

\author{\IEEEauthorblockN{Calvin Deutschbein}
\IEEEauthorblockA{\textit{School of Computing and Information Sciences} \\
\textit{Willamette University}\\
Oregon, United States of America \\
0000-0003-1354-7200}
\and
\IEEEauthorblockN{Aristotle Stassinopoulos}
\IEEEauthorblockA{\textit{School of Computing and Information Sciences} \\
\textit{Willamette University}\\
Oregon, United States of America \\
0009-0008-9913-3140}
}

\maketitle
\conf{\textit{  V. International Conference on Electrical, Computer and Energy Technologies (ICECET 2025) \\ 
3-6 July 2025, Paris-France}}
\begin{abstract} Addressing TedX, Amber Huffman~\cite{huffman24} made an impassioned case that ``none of us is as smart as all of us" and that open-source hardware is the future. A major contribution to software quality, open-source and otherwise, on the software side, is the systems design methodology of Continuous Integration and Delivery (CI/CD), which we propose to systematically bring to hardware designs and their specifications. To do so, we automatically generate specifications using specification mining, ``a machine learning approach to discovering formal specifications"~\cite{ammons02} which dramatically impacted the ability of software engineers to achieve quality, verification, and security. Yet applying the same techniques to hardware is non-trivial. We present a technique for generalized, continuous integration (CI) of hardware specification designs that continually deploys (CD) a hardware specification. As a proof-of-concept, we demonstrate Myrtha, a cloud-based, specification generator based on established hardware and software quality tools.
\end{abstract}


\begin{IEEEkeywords}
Hardware, Security, Machine learning, Cloud computing, RISC-V, Open source, Containers, CI/CD, RTL, IaC, Specification Mining, Formal Verification.
\end{IEEEkeywords}

\section{Introduction}

Continuous Integration (CI) was first proposed in 1991 by Grady Booch for the software domain~\cite{booch91} as a once-per-day, automated, integration test. Since then, CI has exploded in popularity, especially as the broader ``CI/CD" (for continuous integration and delivery) framework, a dominant framework for software quality assurance in recent years. In 2018, the launch of cloud-based solution ``GitHub Actions"~\cite{saroar23}, by GitHub, the host of the Linux Kernel, Python, and TensorFlow, led to a surge of CI/CD. But what about hardware?

Modern hardware designs are exceedingly complex~\cite{schoeberl23}, on the order of 10 billion transistors for consumer CPUs (central processing unit). To combat complexity, Hardware Description Languages (HDLs) enable hardware designers to design software by writing code. Many established language-based software tools may be adapted to HDL. Hardware trends have followed e.g. open-source trends~\cite{huffman24} and formal verification trends including specification mining~\cite{ammons02}.

Many RISC-V~\cite{asanovic14} CPU designs are maintained, like software, under version control on GitHub, increasingly under automated testing. Despite this, we are unaware of a continuous deployment framework for hardware, which is unsurprising for currently existing physical devices. The product of hardware design, however, is not only a physical device, but also a specification document that can be directly interpreted by clients of hardware designs, such as compiler designers, embedded systems engineers, or security researchers.

In this work, we will demonstrate how to systematically apply CI, CD, and specification mining to hardware designs. We organize this around the inversion of the ``Build, Test, Deploy" framework for CI/CD pipelines. For hardware, as we are deploying a specification generated through a testing process, we invert the first terms to ``Test, Build, Deploy", and use specification mining as the build process, with standard CI and CD technologies.
We perform all steps containerized on the cloud, for scalability and transparency. We recognize that a simulation-only approach is insufficient for some hardware goals but still supports hardware quality assurance.

\begin{enumerate}
    \item \textbf{Test}: Using established hardware tools and an existing testbench, we simulate a hardware design to generate a trace of execution as part of CI.
    \item \textbf{Build}: Using specification mining, we build a design specification from the trace data.
    \item \textbf{Deploy}: Using GitHub Actions for CD, we finally deliver the specification as a build artifact.    
\end{enumerate}

\section{Methodology}

\begin{figure}[htbp]
\centerline{\includegraphics[width=0.5\textwidth]{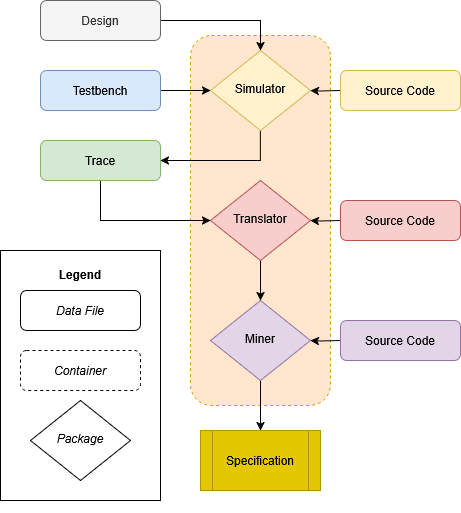}}
\caption{A graph representation of the pipeline}
\label{graph}
\end{figure}

We organize our methodology around managing primary (hardware design) and secondary (software) inputs and encapsulating to manage complexity.

\subsection{Hardware Requirements}

\subsubsection{A Design}

One or more HDL files.

The primary input is a hardware design specified in an HDL such as Verilog or VHDL. In general, we expect a design specified at the register transfer level (RTL).

\subsubsection{A Testbench}

One or more HDL files.

An HDL description of hardware cannot be executed and therefore cannot generate a trace of execution, which is necessary for specification mining. So, we introduce the additional requirement of a testbench. Testbench generation is a separate and active area of research~\cite{zheng24}, but we only require that some imperatives be dispatched to a hardware design from a simulation framework that may log the hardware state.

In general, we find that testbenches are often maintained under version control in the same HDL as the design to which they accompany, as development without testbenches is exceedingly difficult and uncommon. For all designs we explored, we were able to use existing testbenches written by project maintainers.

\subsubsection{A Simulator}

Source code for some software that simulates hardware.

To generate a trace of execution, we must execute hardware in simulation (or generate an equivalent design with hardware monitors, a separate, active area of research~\cite{singh24}). For a cloud-based and scalable solution, we instead use simulation, which has limitations with respect to hardware design but is suitable for the generation of specifications.

In our experience, we found the best approach to simulation was via encapsulation, specifically containerization. We built our simulator from the source within a container and then removed the source code to reduce memory footprint. This led to lightweight, powerful containers with no external dependencies that could be easily deployed to cloud services, and completely abstracted the complexity of hardware simulation from our workflow.

\subsection{Specification Mining Requirements}

\subsubsection{A Trace}

An intermediate data file of unspecified type.

The hardware ``test" phase terminates with the generation of a hardware trace of execution. In practice, these are often ``value change dump" or .vcd files, which specify all changes to the internal state of a hardware design while executing some series of imperatives.

\subsubsection{A Translator}

A custom executable or script.

To our knowledge, there is no widely-used, general-purpose specification miner that accepts traces of execution from software designs, so we implemented our own translation framework. Our framework transposes traces of hardware execution into a format consistent with traces produced by software for software specification miners. In practice, we translated from the .vcd format to a trace format for a C or C++ executable. The executable trace format was similar to HDLs in terms of data types.

\subsubsection{A Miner}

Source code for software that implements the specification mining machine learning process.

To generate a specification from a trace of execution, specification miners infer some universe of candidate properties and then systematically falsify candidate properties while traversing a trace(s) of execution. This process is parallelizable on many axes and may scale quite well, even on larger designs. While arbitrarily sophisticated machine learning techniques may be employed, the specification miners we surveyed tended to rely on heuristic-based algorithms, like \textit{k}-means or hierarchical clustering.

As with the simulator, miners are often large and sophisticated pieces of software with complex build processes. In our case, we converged on a specification miner with a Java Virtual Machine (JVM) dependency, but otherwise maintained only a single Java ARchive, or .jar, file within the container.

\subsection{The Pipeline}

We present the pipeline visually in Fig.~\ref{graph}.  Consistent with the practice of ``Infrastructure as Code (IaC)", all software components implementing the pipeline are built into a single container image. This image contains a hardware simulator, a translator, and a specification miner. While our image also contains the JVM and Python runtimes, which we leveraged for our translator, any pipeline stages could instead be implemented as binaries, as was the hardware simulator.

This container image then becomes a single, separately maintained dependency for both the hardware ``test" stage—which produces a trace—and the software ``build" stage—which produces a specification. We then simply provide the remaining inputs - the hardware design and the testbench - to this container and execute brief script that generates and deploys the specification, in our case as a single .yml (variously ``yet another markup language" or ``yaml ain't markup language") file following GitHub Action standards in order to ``deploy".

We had one remaining input not covered here: in all cases, we additionally used a Makefile to generate traces. Many hardware designs already provided Makefiles which we were able to easily adapt to our workflows.

\section{Implementation}

We implement our methodology with ``Mythra", an open-source package as a container image maintained publicly on GitHub, itself under CI/CD via GitHub actions to GitHub Container Registry (GHCR)~\footnote{\url{https://github.com/hwcicd/myrtha/pkgs/container/myrtha}}. 

\subsection{Simulator: Icarus Verilog}

Icarus Verilog is a free and open-source Verilog compiler under the GPL license, maintained on GitHub~\footnote{\url{https://github.com/steveicarus/iverilog}}. ``Icarus Verilog is not aimed at being a simulator in the traditional sense, but a compiler that generates code employed by back-end tools." In our case, Icarus Verilog is suitable for creating a .vcd file given some Verilog input.

Both Verilog and VHDL are popular for hardware designs, but most designs we were aware of used Verilog (or SystemVerilog). Our toolchain should be fully compatible with other compilers, like Verilator or commercial tools.

Both Verilator and Icarus Verilog use the .vcd text-based format, incurring write-speed limits at one-eighth of the speed of a binary representation (when printing textual binary, one bit of information incurs eight bits of storage). Separately, there is no compelling reason to store the .vcd representation at all, versus streaming directly into a machine learning framework. We hope to approach data streaming in future research and have already adapted our custom translator for streaming data.

\subsection{Translator: Custom Python}

We are aware of no specification miner for .vcd files, and prefer to use established miners as a proof-of-concept regardless. So we required an intermediate stage to translate from the .vcd format to some format suitable for input into our miner. Our miner required ``.decls" declaration files, enumerating variables, and ``.dtrace" Daikon trace files, enumerating the values of all variables at every time point. We note that these files are also text-based formats with corresponding speed limitations.

We implemented this functionality with a Python script that streamed text data file-to-file while maintaining an internal state only large enough to track a current value for each register. Our performance metrics prior to the use of streaming suggested a performance bottleneck in the translation stage due to memory footprint. With the streaming implementation, we shift our bottleneck to Daikon, our externally maintained specification miner.

We recognize that including a full Python installation in a container to use a single script is an inefficient use of container size. In future work, we hope to refactor our script into an executable. Alternatively, we may wish to preserve the Python installation and leverage Python APIs for GPU acceleration in the machine learning stage, in which case the primary change would be to use PyArrow or PyTorch data structures rather than Python built-ins.

\subsection{Miner: The Daikon invariant detector}

``Daikon~\cite{ernst07} is an implementation of dynamic detection of likely invariants; that is, the Daikon invariant detector reports likely program invariants. An invariant is a property that holds at a certain point or points in a program; these are often seen in assert statements, documentation, and formal specifications... Daikon can detect properties in C, C++, and other data sources. (Dynamic invariant detection is a machine learning technique that can be applied to arbitrary data.) It is easy to extend Daikon to other applications." We agree.

We regard specification mining as a separate, ongoing area of research~\cite{lemieux15} and elected to simply use Daikon in our pipeline given our understanding of its popularity. Daikon is implemented as .jar file. Its C/C++ language front-end ``Kvasir" is well-suited to model HDL registers of fixed-width binary values. We made only one specification-relevant design decision when translating .vcd files to Daikon traces, which was how to regard the ``x" unknown and ``z" high impedance values that exist at a hardware level but do not persist at the software level. By treating binary data as unsigned values, we can use the sign bit to indicate hardware-specific values. This results in minor information loss conflating ``x" and ``z", and may require wider words for some values, but was suitable for our purposes.

Consistent with the Daikon documentation, we chose to use a packaged release instead of building from the source. 

In our current pipeline, this stage has by far the highest time cost: a Daikon must read all of the trace data in a text-based format before processing. Daikon usage is motivated by a desire for consistency with existing tools, but we note Daikon uses algorithms for \textit{k}-means or hierarchical clustering that could proceed under GPU-accelerated binary data, possible streaming data, through PyTorch or CUDA C++, and any pipeline will likely have a C/C++ dependency for its hardware simulator. We regard this as an area of future work.

\subsection{The Pipeline}

\begin{figure}[htbp]
\centerline{\includegraphics[width=0.5\textwidth]{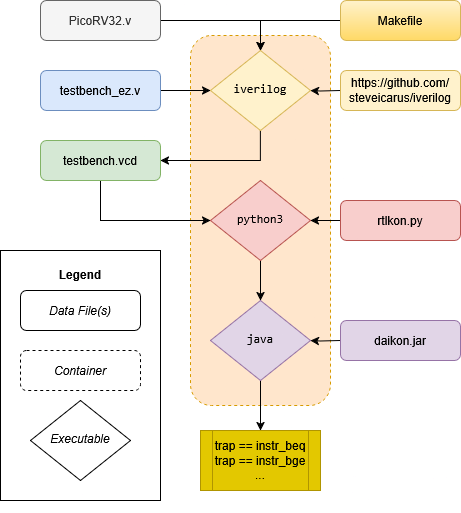}}
\caption{A graph representation the Mythra implementation}
\label{impl}
\end{figure}

We present the pipeline visually in Fig.~\ref{impl}.  It is identical to the proposed methodology except for 3 changes:
\begin{enumerate}
    \item A Makefile is used in .vcd generation.
    \item The translator is a Python script, rather than an executable.
    \item The miner is a JVM package, rather than an executable.
\end{enumerate}

\section{Evaluation}

We developed our implementation over PicoRV32~\footnote{\url{https://github.com/YosysHQ/picorv32}}, ``a CPU core that implements the RISC-V RV32IMC Instruction Set", for which we can report design size, lines of code (LoC), execution times, and specification output. Some evaluations are public via the HWCICD organization~\footnote{\url{https://github.com/hwcicd}}.

\subsection{PicoRV32}

PicoRV32 contains 232 registers in 3049 lines of Verilog code. Its accompanying testbench is 86 lines of Verilog and runs for 2201 cycles. Our performance bottleneck, reading the .dtrace file, scales with the product of unique registers times clock cycles, and is most visible by observing the disk size of the .vcd, .decls, and .dtrace file. The .vcd  also scales with this product, but non-linearly due to tracking on value changes, rather than clock cycles. The .decls scales only with design size. We present these size measures in Tab.~\ref{tab:size}.

\begin{table}[h]
\centering
\caption{PicoRV32 trace data}
\label{tab:size}
\begin{tabular}{lrrr}
& lines & words & bytes  \\
.vcd & 30356 & 46200 & 269184 \\
.decls & 2219 & 4434 & 39059 \\
.dtrace & 2936134 & 2931732 & 17474090 \\
\end{tabular}
\end{table}

\subsection{Myrtha}

\begin{figure}[htbp]
\centerline{\includegraphics[width=0.5\textwidth]{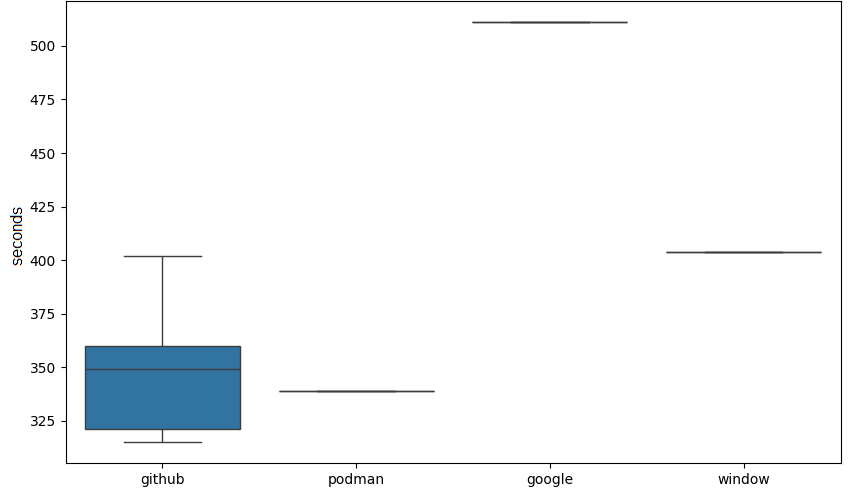}}
\caption{Myrtha build times by platform.}
\label{builds}
\end{figure}

Mythra is implemented as a package, structured over a total of 144 LoC across Python (71), a Containerfile (33), GitHub .yml workflow (32) , and Makefile (8). In general, it takes on the order of 5 minutes or 300 seconds to build the package. Over ten runs on GitHub, Myrtha built in on average in 345 seconds (349 median) with a standard deviation of 27 seconds. We built locally on a Linux device in 339 seconds. Using Docker instead of Podman, we built on Google Cloud Platform (GCP) in 511 seconds and on a local Windows device in 404 seconds. Build times were mostly dominated by compiling Icarus Verilog from source, which required both a lengthy compilation process and the download of a number of packages. These times are summarized in Fig.~\ref{builds}.

Myrtha is a 1.72 GB image over an Ubuntu base. In future efforts, we hope to shift to an Alpine base and minimize the usage of Python, the JVM, and elements of Ubuntu build-essentials not needed by hardware compilers to achieve a lower memory footprint. In practice, the Myrtha container size is reasonable for our application and did not constitute a performance bottleneck.

\subsection{CI/CD}

To perform CI/CD via Myrtha over PicoRV32, we forked the main PicoRV32 repository to our GitHub organization and updated the Makefile and workflow. In total, we added 3 lines to the Makefile generating a specification and developed a 16-line ``myrtha.yml" workflow. GitHub actions ran an average of 39.1 seconds (38 median) with a standard deviation of 3.3 seconds. We expect these times are dominated by initialization, as the workflow runs on the default ubuntu:latest virtual machine hosting the Mythra image. Running locally without a VM, the time to complete the workflow ranged from 4 to 4.5 seconds.

In future research, we hope to generalize our method to other CI/CD platforms. GitLab, for instance, maintains both a cloud-hosted and a community edition that may be hosted locally, so we can gain more detailed performance metrics over what drives costs during the CI/CD hardware stages.

\subsection{Specification}

The output specification is a 4059-line text file summarizing binary equality and inequality operations between registers, modular relations between registers, and linear combinations of registers. This output is suitable to be regarded as a specification of the agreements maintained by PicoRV32, or as an input to a more mature specification generation, such as the security specification generators that derive security agreements from this form of output ~\cite{deutschbein18,deutschbein20,deutschbein21}.  We present a few example invariants in Fig.~\ref{spec}.
\begin{figure}
\begin{verbatim}
4.294967283E9 * dbg_valid_insn + 
  decoded_imm - 4.294967284E9 * 
  is_lui_auipc_jal - 4.294967283E9 == 0
mem_wordsize % q_insn_opcode == 0
trap == eoi
\end{verbatim}
\caption{Example Specifications}
\label{spec}
\end{figure}

\subsection{Evaluation over Holdout Designs}

After developing Myrtha alongside PicoRV32, we tested our approach over other designs. We applied the pipeline to AKER~\footnote{\url{https://github.com/KastnerRG/AKER-Access-Control}}, a design and verification framework for SoC access control~\cite{restuccia21}, and NERV, ``a very simple single-stage RV32I processor.\footnote{\url{https://github.com/YosysHQ/nerv}}". We use a testbench for AKER authored by Andres Meza, with our thanks.

\subsubsection{AKER}

With two changes, the entire pipeline ran without issue on the first attempt. To apply Myrtha to AKER, we used the exact Makefile and Myrtha.yml file used with PicoRV32, updating two lines of code, both within the Makefile:

\begin{enumerate}
    \item \textbf{testbench.vcd target}: We updated the dependencies to refer to the AKER modules.
    \item \textbf{testbench.vcd rule}: We added the -g2012 flag to iverilog to compile the .sv SystemVerilog testbench.  
\end{enumerate}
AKER contains 432 registers in two files totaling 2002 Verilog LoC. Its accompanying testbench is 527 lines of SystemVerilog and runs for 1055 cycles. We present these size measures in Tab.~\ref{tab:aker}. When compared PicoRV32, as predicted, the .decls roughly doubled in size due to the doubling number of registers, but the .dtrace file remained roughly the same size as the clock cycles halved, preserving the product. With similar trace size, we have a similar time cost in GitHub actions, with an average of 41.7 seconds (median 41) and a standard deviation of 3.1 seconds. 

\begin{table}[h]
\centering
\caption{AKER trace data}
\label{tab:aker}
\begin{tabular}{lrrr}
& lines & words & bytes  \\
.vcd & 6290 & 10990 & 53007 \\
.decls & 3519 & 7034 & 62829 \\
.dtrace & 2230270 & 2228160 & 13840288 \\
\end{tabular}
\end{table}

Local times always ranged from 6 to 6.5 seconds, a roughly one-third time increase consistent with a .decls read bottleneck as the size of the .decls file increased by roughly one-third. The spec was 5580 lines, unsurprisingly larger given the doubling of registers but not scaling linearly with design size. Separately, longer traces tend to have fewer properties (as they are falsified over time) and the AKER testbench is shorter. Collectively, these metrics are a positive indicator of our scalability. Additionally, AKER required no changes to Myrtha, so there no marginal cost to pipeline management for placing these additional designs under CI/CD.

\subsubsection{NERV}

To apply Myrtha to NERV, we used the exact Makefile and Myrtha.yml file used with PicoRV32, updating two lines of code, both within the Makefile:

\begin{enumerate}
    \item \textbf{testbench.vcd target}: We updated the dependencies to refer to the NERV modules.
    \item \textbf{testbench.vcd rule}: We adapted the testbench rule from the NERV repository.
\end{enumerate}

We encountered one bug when evaluating NERV: some SystemVerilog features used by NERV were re-categorized from warnings to errors by the latest release of Icarus Verilog. We rolled back from v13 to v11, an earlier stable branch, by changing one line of code in the Myrtha Containerfile and rebuilding. An earlier version of Myrtha had already used v11, but we had previously switched to the main branch to reduce LoC. Once we reverted to the older version, the pipeline ran without issue and we were able to verify that the Myrtha container with the older version of Icarus Verilog was suitable for PicoRV32 and AKER as well.

NERV contains 549 registers in 1267 lines of SystemVerilog. Its accompanying testbench is 155 SystemVerilog LoC and runs for 19 cycles. We present these size measures in Tab.~\ref{tab:nerv}. GitHub actions run in an average of 36.8 seconds (median 36) with a standard deviation of 4 seconds. 

\begin{table}[h]
\centering
\caption{NERV trace data}
\label{tab:nerv}
\begin{tabular}{lrrr}
& lines & words & bytes  \\
.vcd & 2248 & 6689 & 34848 \\
.decls & 5379 & 10754 & 101427 \\
.dtrace & 61370 & 61332 & 483308 \\
\end{tabular}
\end{table}

\section{Related Work}

\paragraph*{Hardware Specification Mining}
Early hardware specification leveraged known patterns, such as one-hot encoding~\cite{hangal05,mandouh12}. Other researchers applied data mining~\cite{chang10,hertz13,li10} or temporal logic~\cite{danese2017team,danese2015automaticdate,danese2015automatic,deutschbein18} techniques. 
Security researchers have manually identified properties~\cite{hicks15,irvine11,brown17} from automated generation for RISC~\cite{ZhangASPLOS2017} and CISC~\cite{deutschbein20} designs and discovered subsets of hyperproperties~\cite{rawat20,deutschbein21}.

\paragraph*{Specification Mining}
Ammons et al. introduced specification mining~\cite{ammons02} and launched a rich research direction across static and dynamic analysis~\cite{weimer05}; imperfect traces ~\cite{yang06}; and complex traces~\cite{gabel2008javert,reger13,gabel2008symbolic}. Perhaps the most widely known miner, Daikon~\cite{ernst07} approached specification mining as invariant detection.

\paragraph*{Hardware CI/CD Pipelines}
Despite widespread adoption in software, to the best of our knowledge there is minimal research on continuous integration for hardware design specification, outside of a few examples of Continuous Integration/Continuous Delivery (CI/CD) approaches to the embedded space~\cite{talekar23, prakasia24}, which explores software and hardware together. By contrast, there is ample research on hardware acceleration for software CI/CD~\cite{shahin17}.

\section{Conclusion}

We have proposed ``Test, Build, Deploy" and demonstrate Myrtha, a proof-of-concept for a CI/CD cloud-based machine learning framework that generalized well to other hardware designs.

\section*{Acknowledgements}

We thank the NSF for funding. We thank Andres Meza for the AKER testbench. We thank  Intel Corporation, the Kastner Research Group, and the HWSec@UNC research group for helpful discussions. We thank our undergraduate researchers, Kendall Leonard and Hannah Pahama, for suggesting the use of containerization and ``ClickOps" to make hardware research more accessible.

\bibliographystyle{unsrt}
\bibliography{refs}

\vspace{12pt}

\end{document}